%Corrections by AEL, 8/29/02
%''Now I'm being INVOLUNTARILY shuffled towards the
%CLAM DIP with the BROKEN PLASTIC FORKS IN IT!''

\let\includefigures=\iftrue
\input psfig
%
% the following is to use blackboard bold fonts --
\let\useblackboard=\iftrue
%
% activate this if you don't have them.

%\let\useblackboard=\iffalse
%
% You might also need to remove this line.
\newfam\black
\input harvmac
\noblackbox
%%
%Figure Stuff
\includefigures
\message{If you do not have epsf.tex (to include figures),}
\message{change the option at the top of the tex file.}
\input epsf
\def\figin{\epsfcheck\figin}\def\figins{\epsfcheck\figins}
\def\epsfcheck{\ifx\epsfbox\UnDeFiNeD
\message{(NO epsf.tex, FIGURES WILL BE IGNORED)}
\gdef\figin##1{\vskip2in}\gdef\figins##1{\hskip.5in}% blank space
instead
\else\message{(FIGURES WILL BE INCLUDED)}%
\gdef\figin##1{##1}\gdef\figins##1{##1}\fi}
\def\DefWarn#1{}
\def\figinsert{\goodbreak\midinsert}
\def\ifig#1#2#3{\DefWarn#1\xdef#1{fig.~\the\figno}
\writedef{#1\leftbracket fig.\noexpand~\the\figno}%
\figinsert\figin{\centerline{#3}}\medskip\centerline{\vbox{
\baselineskip12pt\advance\hsize by -1truein
\noindent\footnotefont{\bf Fig.~\the\figno:} #2}}
\bigskip\endinsert\global\advance\figno by1}
%%%
\else
\def\ifig#1#2#3{\xdef#1{fig.~\the\figno}
\writedef{#1\leftbracket fig.\noexpand~\the\figno}%
%\figinsert\figin{\centerline{#3}}\medskip
%\centerline{\vbox{\baselineskip12pt
%\advance\hsize by -1truein\noindent
%\footnotefont{\bf Fig.~\the\figno:} #2}}
%\bigskip\endinsert
\global\advance\figno by1}
\fi
%
%%this is to put a figure without a caption

%%this is to put a figure without whitespace afterward
\def\smallfig#1#2#3{\DefWarn#1\xdef#1{fig.~\the\figno}
\writedef{#1\leftbracket fig.\noexpand~\the\figno}%
\figinsert\
in{\centerline{#3}}\medskip\centerline{\vbox{
\baselineskip12pt\advance\hsize by -1truein
\noindent\footnotefont{\bf Fig.~\the\figno:} #2}}
\endinsert\global\advance\figno by1}

%%BLACKBOARD FONT STUFF
\useblackboard
\message{If you do not have msbm (blackboard bold) fonts,}
\message{change the option at the top of the tex file.}
\font\blackboard=msbm10 scaled \magstep1
\font\blackboards=msbm7
\font\blackboardss=msbm5
\textfont\black=\blackboard
\scriptfont\black=\blackboards
\scriptscriptfont\black=\blackboardss

\else

\fi
% *************************************
%\draft
%

% Macros for boxes from cordes moore ramgoolam

\def\boxit#1{\vbox{\hrule\hbox{\vrule\kern8pt
\vbox{\hbox{\kern8pt}\hbox{\vbox{#1}}\hbox{\kern8pt}}
\kern8pt\vrule}\hrule}}
\def\mathboxit#1{\vbox{\hrule\hbox{\vrule\kern8pt\vbox{\kern8pt
\hbox{$\displaystyle #1$}\kern8pt}\kern8pt\vrule}\hrule}}

%% for sub sub sections
\def\subsubsection#1{\bigskip\noindent
{\it #1}}

\def\yboxit#1#2{\vbox{\hrule height #1 \hbox{\vrule width #1
\vbox{#2}\vrule width #1 }\hrule height #1 }}
\def\fillbox#1{\hbox to #1{\vbox to #1{\vfil}\hfil}}
\def\ybox{{\lower 1.3pt \yboxit{0.4pt}{\fillbox{8pt}}\hskip-0.2pt}}
%
%

%%MATH MACROS
%Greek letters and their bars

%More bars

\def\l{\left}

\def\comments#1{}

\def\p{\partial}

\def\half{{1\over 2}}

\def\Re{{\rm Re\hskip0.1em}}
\def\Im{{\rm Im\hskip0.1em}}

\def\bra#1{{\langle}#1|}
\def\ket#1{|#1\rangle}

%AEL

\def\CL{{\cal L}}

%AEL

\def\II{\relax{I\kern-.10em I}}

\font\cmss=cmss10 \font\cmsss=cmss10 at 7pt
\def\IZ{\relax\ifmmode\mathchoice
{\hbox{\cmss Z\kern-.4em Z}}{\hbox{\cmss Z\kern-.4em Z}}
{\lower.9pt\hbox{\cmsss Z\kern-.4em Z}}
{\lower1.2pt\hbox{\cmsss Z\kern-.4em Z}}
\else{\cmss Z\kern-.4emZ}\fi}
\def\IR{\relax{\rm I\kern-.18em R}}
\def\IZ{\relax\ifmmode\mathchoice
{\hbox{\cmss Z\kern-.4em Z}}{\hbox{\cmss Z\kern-.4em Z}}
{\lower.9pt\hbox{\cmsss Z\kern-.4em Z}} {\lower1.2pt\hbox{\cmsss
Z\kern-.4em Z}}\else{\cmss Z\kern-.4em Z}\fi}
\def\IB{\relax{\rm I\kern-.18em B}}
\def\IC{{\relax\hbox{$\inbar\kern-.3em{\rm C}$}}}
\def\ID{\relax{\rm I\kern-.18em D}}
\def\IE{\relax{\rm I\kern-.18em E}}
\def\IF{\relax{\rm I\kern-.18em F}}
\def\IG{\relax\hbox{$\inbar\kern-.3em{\rm G}$}}
\def\IGa{\relax\hbox{${\rm I}\kern-.18em\Gamma$}}
\def\IH{\relax{\rm I\kern-.18em H}}
\def\II{\relax{\rm I\kern-.18em I}}
\def\IK{\relax{\rm I\kern-.18em K}}
\def\IP{\relax{\rm I\kern-.18em P}}
%\def\IX{\relax{\rm X\kern-.01em X}}
%this doesn't work

%

\def\inbar{\,\vrule height1.5ex width.4pt depth0pt}

\def\p{\partial}

\font\cmss=cmss10 
\def\IR{\relax{\rm I\kern-.18em R}}

%

 % for now

%

\def\lp10{\ell_p^{10}}
\def\lp11{\ell_p^{11}}
\def\R11{R_{11}}

\def\frac#1#2{{#1 \over #2}}

%More bars

\def\l{\left}

\def\comments#1{}

\def\p{\partial}

\def\half{{1\over 2}}

\def\Re{{\rm Re\hskip0.1em}}
\def\Im{{\rm Im\hskip0.1em}}

\def\bra#1{{\langle}#1|}
\def\ket#1{|#1\rangle}

%AEL

\def\CL{{\cal L}}

%AEL

%% New Defs
\def\Hor{Ho\v{r}ava}
\def\ibid{{\it ibid.}}
\def\cf{{\it c.f.}}
\def\M4{M_{Pl,4}}

\def\k11{\kappa_{11}}
\def\l11{\ell_{11}}
\def\tl11{\tilde{\ell}_{11}}

\def\m11{M_{11}}
\def\tm11{\tilde{M}_{11}}

%% REF MACROS
\def\np{{ Nucl. Phys.}}
\def\prl{{ Phys. Rev. Lett.}}
\def\pr{{ Phys. Rev.}}
\def\pl{{ Phys. Lett.}}

\def\anpny{{ Ann. Phys.}\ (N.Y.)}

%% Sec 1 Refs
\lref\kolbturner{E.W. Kolb and M.S. Turner, {\it The Early Universe}
Addison-Wesley (1990).}
\lref\lindebook{A. Linde,
{\it Particle Physics and Inflationary Cosmology}
Harwood Academic, (1990).}

\lref\liddlelyth{A. Liddle and D. Lyth, {\it Cosmological
Inflation and Large-Scale Structure} Cambridge University Press
(2000).}

\lref\reconsrev{For a review see
J. E. Lidsey {\it et. al.},
``Reconstructing the Inflaton Potential--An Overview,"
Reviews of Modern Physics, {\bf 69} (1997) 373.}
\lref\lindehybrid{A. Linde, ``Axions in Inflationary Cosmology,"
\pl {\bf B259} (1991) 38; {\it ibid.},
``Hybrid Inflation," \pr {\bf D49} (1994) 748, astro-ph/9307002.}

\lref\naturalinfl{K. Freese, J.A. Frieman and A.V. Olinto,
``Natural Inflation with Pseudo-Nambu-Goldstone Bosons",
Phys. Rev. Lett. {\bf 65} (1990) 3233;
F.C. Adams, J.R. Bond, K. Freese, J.A. Frieman and A.V. Olinto,
``Natural Inflation:  Particle Physics Models,
Power Law Spectra for Large Scale Structure, and Constraints from
COBE'', Phys. Rev. {\bf D47} (1993) 426, hep-ph/9207245}

\lref\wimpz{D.J.H. Chung, E.W. Kolb, A. Riotto and I. Tkachev,
``Probing Planckian Physics: Resonant Production of Particles
During Inflation and Features in the Primordial Power Spectrum,''
Phys.Rev. {\bf D62} (2000) 043508, hep-ph/9910437.}

\lref\martinbrand{J. Martin and R. H. Brandenberger,
``The Transplanckian Problem of Inflationary
Cosmology," Phys. Rev. {\bf D63} (2001) 123501, hep-th/0005209; R. H.
Branbenberger and J. Martin,
``The Robustness of Inflation to Changes
in Superplanck Scale Physics," Mod. Phys. Lett. {\bf A16} (2001)
999,  astro-ph/0005432; R. H. Brandenberger, S. E. Joras and J.
Martin,
Trans-Planckian Physics and the Spectrum
of Fluctuations in a Bouncing Universe, hep-th/0112122 . }

\lref\Niem{J. C. Niemeyer, ``Inflation With a Planck Scale
Frequency Cutoff," Phys. Rev. {\bf D63} (2001) 123502,
astro-ph/0005533; J. C. Niemeyer and R. Parentani,
``Transplanckian Dispersion and Scale invariance
of Inflationary Perturbations,"
Phys. Rev. {\bf D64} (2001) 101301,  astro-ph/0101451.}

\lref\kempf{A. Kempf and J. C. Niemeyer,
``Perturbation Spectrum in Inflation With Cutoff,"
Phys. Rev. {\bf D64} (2001) 103501,
astro-ph/0103225.}

\lref\kolbresonant{D.J.H. Chung, E.W. Kolb, A. Riotto and I.
Tkachev, ``Probing Planckian Physics: Resonant Production of
Particles During Inflation and Features in the Primordial Power
Spectrum," Phys. Rev. {\bf D62} (2000) 043508, hep-ph/9910437.}

\lref\stretched{L. Susskind, L. Thorlacius and J. Uglum,
``The stretched horizon and black hole complementarity,''
Phys. Rev. {\bf D48} (1993) 3743,
hep-th/9306069.}
%%CITATION = HEP-TH 9306069;%%}

\lref\desreview{M. Spradlin, A.
Strominger and A. Volovich, ``Les Houches lectures on de Sitter
space,'' hep-th/0110007.}
\lref\hawkingellis{S. Hawking and G.F.R.
Ellis, {\it The large-scale structure of spacetime}, Cambridge
(UK).}
%\lref\liddlelyth{Liddle and Lyth, inflation book}
\lref\birrelldavies{N.D. Birrell and P.C.W. Davies, {\it Quantum
fields in curved space}, Cambridge Univ. Press (1982), Cambridge,
UK.}
\lref\waldbook{R.M. Wald, {\it Quantum Field Theory in Curved
Spacetime and Black Hole Thermodynamics}, U. Chicago Press
(1994).}

\lref\egks{R. Easther, B. R. Greene, W. H. Kinney,
and G. Shiu, ``Inflation as a Probe
of Short Distance Physics, " Phys. Rev. {\bf D64} (2001) 103502,
hep-th/01044102.}

\lref\egkss{R. Easther, B. R. Greene,
W. H. Kinney, and G. Shiu, ``Imprints of Short Distance
Physics on Inflationary Cosmology," hep-th/0110226 .}

\lref\witstrong{E. Witten, ``Strong Coupling
Expansion of Calabi-Yau Compactification," \np\ {\bf B471} (1996) 135,
hep-th/9602070}

\lref\conscon{L. Hui and W. H. Kinney,
``Short Distance Physics and the Consistency Relation
for Scalar and Tensor Fluctuations in the Inflationary Universe,"
astro-ph/0109107.  For examples of other models that discuss
modifications of the consistency conditions, see {\it e.g.},
J. Garriga and V. F. Mukhanov, ``Perturbations in K-Inflation,"
Phys. Lett. {\bf B458} (1999) 219, hep-th/9904176,
G. Shiu and S. H. H. Tye, ``Some Aspects of Brane Inflation,"
Phys. Lett. {\bf B516} (2001) 421, hep-th/0106274.
}

\lref\wang{L. Wang, V. Mukhanov, and P. Steinhardt, ``On the Problem
of
Predicting Inflationary Perturbations," Phys. Lett. {\bf B414} (1997)
18,
astro-ph/9709032.}

\lref\mukh{V. Mukhanov and G. Chibisov, ``Quantum Fluctuations and
a Nonsingular Universe", JETP Lett. {\bf 33} (1981) 532.}

\lref\bst{J. Bardeen, P. Steinhardt, and M. Turner, ``Spontaneous
Creation of Almost Scale-Free Density Perturbations in an Inflationary
Universe," Phys. Rev {\bf D28} (1983) 679.}

\lref\hawk{S. W. Hawking, ``The Development of Irregularities in a
SIngle
Bubble Inflationary Universe", Phys. Lett. {\bf B115} (1982) 295.}

\lref\guth{A. H. Guth and S. Y. Pi, ``Fluctuations in the New
Inflationary
Universe", Phys. Rev. Lett. {\bf 49} (1982) 1110.}

\lref\starob{A. A. Starobinsky, ``Dynamics of Phase Transition in the
New
Inflationary Universe Scenario and Generation of Perturbations", Phys.
Lett. {\bf B117} (1982) 175.}

\lref\alinsf{A. Linde, ``Scalar Field Fluctuations in Expanding
Universe
and the New Inlfationary Universe Scneario", Phys. Lett. {\bf B116}
(1982) 335.}

\lref\cgkl{E. Copeland, I. Girvell, E. Kolb, and A. Liddle, ``On the
Reliability of Inflaton
Potential Reconstruction," Phys. Rev. {\bf D58} (1998) 043002,
astro-ph/9802209.}

\lref\cobe{J. C. Mather {\it et al.}, ``A Preliminary Measurement of
the
Cosmic Microwave
Background Spectrum by the Cosmic Microwave Background Explorer (COBE)
Satellite,"
Astrophys. J. {\bf 354} (1990) L37.}

\lref\boomerang{A. E. Lange {\it et al.}, ``Cosmological Parameters
from
the First Results
of BOOMERANG," Phys. Rev. {\bf D63} (2001) 042001, astro-ph/0005004.}

\lref\maxima{Balbi, A. {\it et al.}, ``Constraints on Cosmological
Parameters from MAXIMA-1",
Astrophys. J. {\bf 545} (2000) L5, astro-ph/0005124.}

%%Sec 3 Refs

\lref\bankscosmo{T. Banks, ``Cosmological Breaking of Supersymmetry?
or Little Lambda Goes Back to the Future 2. ," hep-th/0007146 .}

\lref\banksb{T. Banks and L. Mannelli, ``De Sitter vacua, renormalization and locality,"
hep-th/0209113.}

\lref\larsen{M. Einhorn and F. Larsen, ``Interacting quantum field theory in de Sitter
vacua," hep-th/0209159.}

\lref\fischler{W. Fischler, ``Taking de Sitter Seriously,"
talk given at {\it Role of Scaling laws in Physics and Biology
(Celebrating the 60th Birthday of Geoffrey West),} Santa Fe, Dec.
2000.}

%% Sec 5 refs
\lref\selzal{U. Seljak and M. Zaldarriaga,
``Signature of Gravity Waves in Polarization of the Microwave
Background,"
{\it Phys.Rev.Lett.} {\bf 78} (1997) 2054.}

%% Refs
\lref\fs{W. Fischler and L. Susskind, ``Holography and Cosmology,''
hep-th/9806039.}
\lref\rb{R. Bousso, ``A Covariant Entropy Conjecture,''
JHEP {\bf 9907} (1999) 004,
hep-th/9905177; {\it ibid.},
``Holography in general space-times,''
JHEP {\bf 9906} (1999) 028,
hep-th/9906022.}
\lref\hks{S. Hellerman, N. Kaloper and L. Susskind,
``String theory and quintessence,''
JHEP {\bf 0106}, (2001) 003,
hep-th/0104180;
W. Fischler, A. Kashani-Poor, R. McNees and S. Paban,
``The Acceleration of the Universe, a Challenge for String Theory",
JHEP {\bf 0107} (2001) 003, hep-th/0104181.}
\lref\Bard{J. Bardeen,
``Gauge Invariant Cosmological Perturbations,''
\pr\ {\bf D22}\ (1980) 1882.}
\lref\horwit{P. \Hor\ and E. Witten,
``Heterotic and Type I string dynamics
form eleven dimensions'', \np\ {\bf B460}\
(1996) 506, hep-th/9510209; \ibid, ``Eleven-dimensional
supergravity on a manifold with boundary'',
\np\ {\bf B475} (1996) 94, hep-th/9603142.}
\lref\add{N. Arkani-Hamed, S. Dimopoulos
and G.R. Dvali, ``The Hierarchy Problem and
New Dimensions at a Millimeter,''
\pl\ {\bf B429} (1998) 263,
hep-ph/9803315; \ibid,
``Phenomenology, Astrophysics and Cosmology
of Theories with Sub-millimeter Dimensions and TeV
Scale Quantum Gravity,''
\pr\ {\bf D59}\ (1999) 086004,
hep-ph/9807344.}
\lref\kalin{N. Kaloper and A. Linde,
``Inflation and Large Internal Dimensions,''
\pr\ {\bf D59} (1999) 101303,
hep-th/9811141.}
\lref\addcos{N. Arkani-Hamed, S. Dimopoulos,
N. Kaloper and J. March-Russell,
``Rapid Asymmetric Inflation and
Early Cosmology in Theories with Sub-millimeter Dimensions,''
\np\ {\bf B567} (2000) 189,
hep-ph/9903224.}

\lref\unify{S. Dimopoulos and H. Georgi, `` Softly Broken
Supersymmetry
and SU(5)", Nucl. Phys. {\bf B192}
(1981) 150; S. Dimopoulos, S. Raby and F. Wilczek, ``Supersymmetry and
the Scale of Unification", Phys. Rev. {\bf
D24} (1981) 1681;
U. Amaldi, W. de Boer and H. Furstenau,
``Comparison of Grand Unified Theories with Electroweak and Strong
Coupling
Constants Measured at LEP",
Phys. Lett. {\bf B260} (1991) 447;
P. Langacker and N. Polonsky,
``Uncertainties in Coupling Constant Unification",
Phys. Rev. {\bf D47} (1993) 4028, hep-ph/9210235.}
\lref\linde{A.D. Linde, ``Chaotic Inflation,''
\pl\ {\bf B129} (1983) 177.}
\lref\gsw{M.B. Green, J. Schwarz and E. Witten,
{\it Superstring Theory}, vols. I and II,
Cambridge Univ. Press (1987).}
\lref\gluino{P. \Hor,
``Gluino Condensation in Strongly Coupled
Heterotic String Theory,''
\pr\ {\bf D54} (1996) 7561,
hep-th/9608019.}
\lref\tomcosm{T. Banks, ``Remarks on M Theoretic
Cosmology,'' hep-th/9906126.}
\lref\tomcosmrev{T. Banks, ``M-theory and cosmology,''
hep-th/9911067.}
\lref\feynman{R.P. Feynman et. al., {\it Feynman
Lectures on Gravitation}}

%%AEL NEW REFS
\lref\kamkos{M. Kamionkowski and A. Kosowsky,
``The Cosmic Microwave Background and Particle Physics,''
{\it Ann.\ Rev.\ Nucl.\ Part.\ Sci.}  {\bf 49} (1999) 77,
astro-ph/9904108.}
\lref\gtwofiber{J.~A.~Harvey, D.~A.~Lowe and A.~Strominger,
``N=1 String Duality,''
\pl\ {\bf B362} (1995) 65,
hep-th/9507168; B.~S.~Acharya,
``N=1 Heterotic/M-theory Duality and Joyce Manifolds,''
\np\ {\bf B475} (1996) 579,
hep-th/9603033.}
\lref\gtwochiral{E.~Witten,
``Anomaly Cancellation on $G_2$ Manifolds,''
hep-th/0108165; B.~S.~Acharya and E.~Witten,
``Chiral Fermions from Manifolds of $G_2$ Holonomy,''
hep-th/0109152; E. Witten, ``Deconstruction,
$G_2$ Holonomy, and Doublet-Triplet
Splitting,'' hep-ph/0201018.}

\lref\syz{A. Strominger, S.-T. Yau and E. Zaslow,
``Mirror Symmetry is T-duality,''
\np\ {\bf B479}\ (1996) 243, hep-th/9606040.}
\lref\addcosmprob{K. Benakli and S. Davidson,
``Baryogenesis in Models with a Low Quantum Gravity Scale,''
\pr\ {\bf D60} (1999) 025004, hep-ph/9810280;
D. Lyth, ``Inflation With TeV-scale Gravity Needs
Supersymmetry,'' \pl\ {\bf B448}\ (1999) 191,
hep-ph/9810320.}

\lref\addss{I. Antoniadis, S. Dimopoulos and G, Dvali,
Scherk-Schwarz paper, and other refs?}

\lref\mirpeskin{E. Mirabelli and M. Peskin, 5d
theories on interval.}

\lref\modular{
P. Binetruy and M.K. Gaillard,
``Candidates for the Inflaton Field in Superstring Models,"
Phys. Rev. D34 (1986) 3069;
L. Randall and S. Thomas ,
``Solving the Cosmological Moduli Problem with Weak Scale Inflation,"
Nucl. Phys. {\bf B449} (1995) 229,
hep-ph/9407248;
T. Banks, M. Berkooz, G. Moore, S. Shenker
and P. Steinhardt, ``Modular cosmology,''
\pr\ {\bf D52} (1995) 3548, hep-th/9503114.
S. Thomas,
``Moduli Inflation from Dynamical Supersymmetry Breaking,"
Phys.Lett. {\bf B351} (1995) 424,
hep-th/9503113.}

\lref\ruts{J. Russo and A. Tseytlin,
``One-loop Four-graviton Amplitude in
Eleven-Dimensional Supergravity,''
\np\ {\bf B508} (1997) 245, hep-th/9707134.}
\lref\hettypeI{J. Polchinski and E. Witten,
``Evidence for Heterotic - Type I String Duality,''
\np\ {\bf B460} (1996) 525, hep-th/9510169.}

\lref\dsscales{M. Dine and N. Seiberg, ``Couplings and
Scales in Superstring Models,'' \prl\ {\bf 55}\
(1985) 366.}
\lref\kscales{V. Kaplunovsky, ``Mass
Scales of the String Unification,''
\prl\ {\bf 55}\ (1985) 1036.}
\lref\multif{D. Polarski and A.A. Starobinsky,
``Structure of Primordial Gravitational
Waves Spectrum in a Double Inflationary
Model,'' \pl\ {\bf B356}\ (1995) 196,
astro-ph/9505125; J. Garcia-Bellido and
D. Wands, \pl\ {\bf D52}\ (1995) 6739;
M. Sasaki and E. Stewart, ``A General Analytic
Formula for the Spectral Index of the Density
Perturbations Produced During Inflation,''
{\it Prog. Theor. Phys.}\ {\bf 95}\ (1996) 71,
astro-ph/9507001.}
\lref\corrconsist{N. Bartolo, S. Matarrese and
A. Riotto, ``Adiabatic and Isocurvature Perturbations
from Inflation: Power Spectra and Consistency
Relations,'' \pr\ {\bf D64} (2001) 123504.}
\lref\isoc{A.D. Linde, ``Generation of
Isothermal Density Perturbations in the
Inflationary Universe,'' \pl\ {\bf B158} (1985) 375;
L.A. Kofman, ``What Initial Perturbations may be
Generated in Inflationary Cosmological Models,''
\pl\ {\bf B173}\ (1986) 400; L.A. Kofman and
A. Linde, ``Generation of Density Perturbations
in Inflationary Cosmology,'' \np\ {\bf B282}\ (1987) 555;
A. Linde and V. Mukhanov, ``Non-Gaussian Isocurvature
Perturbations from Inflation,'' \pr {\bf D56} (1997) R535;
P.J.E. Peebles, astro-ph/9805194.}
\lref\efbond{G. Efstathiou and J.R. Bond,
{\it Mon. Not. R. Astron. Soc.} {\bf 218}\
(1986) 103.}
\lref\langlois{D. Langlois, ``Correlated Adiabatic and
Isocurvature Perturbations from Double Inflation,''
\pr\ {\bf D59} (1999) 123512; D. Langlois and
A. Riazuelo, ``Correlated Mixtures of Adiabatic and
Isocurvature Cosmological Perturbations,''
\pr\ {\bf D62}\ (2000) 043504; C. Gordon {\it et. al.},
``Adiabatic and Entropy Perturbations from Inflation,''
\pr\ {\bf D63} (2000) 023506; N. Bartolo, S. Mattarese
and A. Riotto, ``Oscillations During Inflation and the
Cosmological Density Perturbations,'' \pr\ {\bf D64}\
(2001) 083514.}
\lref\correxp{M. Bucher, K. Moodley and N. Turok,
``General Primordial Cosmic Perturbation,'' \pr\ {\bf D62}
(2000) 083508; R. Trotta, A. Riazuelo and R. Durrer,
``Cosmic Microwave Background Anisotropies with
Mixed Isocurvature Perturbations,'' astro-ph/0104107;
L. Amendola {\it et. al.}, ``Correlated Perturbations
{}From Inflation and the Cosmic Microwave Background,''
astro-ph/0107089.}
\lref\huandwhite{For a pedagogical review see
W.~Hu and M.~J.~White,
``A CMB Polarization Primer,''
{\it New Astron.}  {\bf 2} (1997) 323,
astro-ph/9706147.}

%some alpha refs

%AEL NEW 6/27
%\lref\stretched{Stretched horizon refs}
%\lref\wimpz{WIMPzilla refs}
\lref\hogan{C.~J.~Hogan, ``Holographic discreteness of
inflationary perturbations,'' Phys.\ Rev.\ {\bf D66} (2002) 023521,
astro-ph/0201020;
%%CITATION = ASTRO-PH 0201020;%%
``Observing quanta on a cosmic scale,'' Fortsch.\ Phys.\  {\bf
50} (2002) 694, astro-ph/0201021.
%%CITATION = ASTRO-PH 0201021;%%
}
\lref\stevescott{M. Kleban, S.H. Shenker and S. Thomas, unpublished.}
\def\eg{{\it e.g.}}

%%%%%%%%%

\Title{\vbox{\baselineskip12pt\hbox{}
\hbox{BRX TH-505} \hbox{SLAC-PUB-9112} \hbox{SU-ITP-02/02} }} {\vbox{
\centerline{Initial conditions for inflation}}}
\smallskip
\centerline{Nemanja Kaloper$^{1,2}$, Matthew Kleban$^{1}$,
Albion Lawrence$^{1,3,4}$,}
\centerline{Stephen Shenker$^{1}$ and Leonard Susskind$^{1}$}
\bigskip
\bigskip
\centerline{$^{1}${Department of Physics,
Stanford University, Stanford, CA 94305}}
\medskip
\centerline{$^{2}${Department of Physics, University of California,
Davis, CA 95616}}
\medskip
\centerline{$^{3}$Brandeis University Physics Department,
MS 057, POB 549110, Waltham, MA 02454\footnote{${}^\dagger$}
{Present and permanent address.}}
\medskip
\centerline{$^{4}${SLAC Theory Group, MS 81, 2575 Sand Hill Road,
Menlo Park, CA 94025}}
\bigskip
\bigskip
\noindent

Free scalar fields in de Sitter space
have a one-parameter family of states
invariant under the de Sitter group,
including the standard thermal vacuum.  We show
that, except for the thermal vacuum, these states are unphysical
when gravitational
interactions are included.  We apply
these observations to the quantum state
of the inflaton, and find that at best, dramatic
fine tuning is required for states other than
the thermal vacuum to lead to observable features
in the CMBR anisotropy.

\medskip
\bigskip

\Date{August 2002}

%Effective action and the CMBR
\lref\kkls{N. Kaloper, M. Kleban, A. Lawrence and S. Shenker,
``Signatures of short distance physics in the cosmic
microwave background,'' hep-th/0201158.}

%New vacua and the CMBR
\lref\danielsson{U.H. Danielsson, ``A note on inflation
and transplanckian physics,'' hep-th/0203198.}
\lref\greeneII{R. Easther, B.R. Greene, W.H. Kinney and
G. Shiu, ``A generic estimate of transplanckian
modifications to the primordial power spectrum during inflation,''
hep-th/0204129.}
\lref\brandenbergerII{R.H. Brandenberger and J. Martin,
``On signatures of short distance physics in the cosmic
microwave background,'' hep-th/0202142.}
\lref\danielssontwo{U.H. Danielsson, ``Inflation,
Holography, and the Choice of Vacuum iin de Sitter Space,''
hep-th/0205227.}

%Experimental refs
\lref\cobe{J. C. Mather {\it et al.},
``A Preliminary Measurement of the Cosmic Microwave
Background Spectrum by the Cosmic Microwave
Background Explorer (COBE) Satellite,"
Astrophys. J. {\bf 354} (1990) L37.}
\lref\boomerang{A. E. Lange {\it et al.}, ``Cosmological
Parameters from the First Results
of BOOMERANG," Phys. Rev. {\bf D63} (2001)
042001, astro-ph/0005004.}
\lref\maxima{Balbi, A. {\it et al.}, ``Constraints on
Cosmological Parameters from MAXIMA-1",
Astrophys. J. {\bf 545} (2000) L5, astro-ph/0005124.}
\lref\great{N.J. Cornish, D. Spergel and C.L. Bennett,
``Journey to the edge of time: the GREAT mission,''
astro-ph/0202001.}
\lref\greatb{N. Seta, S. Kawamura, and T. Nakamura,
``Possibility of direct measurement of the acceleration
of the universe using 0.1 Hz band laser interferometer
gravitational wave antenna in space," Phy. Rev. Lett {\bf 87}
(2001), 221103-1.}
\lref\greatc{C. Ungarelli, A. Vecchio, ``High energy physics
and the very early universe with LISA," Phys. Rev. D {\bf 63}
(2001) 064030-1.}

%New vacua
\lref\mottola{E. Mottola, ``Particle creation
in de Sitter space,'' \pr\ {\bf D31} (1985) 754.}
\lref\allen{B. Allen, ``Vacuum states in de Sitter space,''
\pr\ {\bf D32} (1985) 3136.}
\lref\cherntag{N.A. Chernikov and E.A. Tagirov,
``Quantum theory of scalar fields in de Sitter spacetime,''
{\it Ann. Inst. Henri Poincar\'e} {\bf A9} (1968) 109;
E.A. Tagirov, ``Consequences of field quantization
in de Sitter type cosmological models,''
\anpny\ {\bf 76}\ (1973) 561.}
\lref\schomblond{J. G\'eh\'eniau and C. Schomblond,
{\it Acad. R. Belg. Bull. Cl. Sci.}\
{\bf 54} (1968) 1147; C. Schomblond and P. Spindel,
{\it Ann. Inst. Henri Poincar\'e} {\bf A25}
(1976)  67.}
\lref\goldlowe{K. Goldstein and D. Lowe,
``Initial state effects on the cosmic microwave
background and trans-planckian physics'',
hep-th/0208167.}

%ds/CFT
\lref\dscft{A. Strominger, ``The dS/CFT correspondence,''
hep-th/0106113.}
\lref\bms{R. Bousso, A. Maloney and A. Strominger,
``Conformal vacua and entropy in de Sitter space,''
hep-th/0112218.}

%Unruh detectors
\lref\unruhd{W.G. Unruh,
``Notes On Black Hole Evaporation,''
Phys.\ Rev.\ {\bf D14} (1976) 870.
%%CITATION = PHRVA,D14,870;%%
}

\lref\mattlennylisa{L. Dyson, M. Kleban, and L. Susskind,
``Disturbing implications of a cosmological constant," hep-th/0208013.
%%CITATION = HEP-TH 0208013;%%
}

%Private discussions
\lref\juanconv{J. Maldacena, private conversation.}

\newsec{Introduction}
%For the introduction
\def\tvev{\langle T_{\mu\nu}\rangle}

In inflationary cosmology, cosmic microwave background (CMB)
data place a tantalizing upper bound on the vacuum
energy density during the
inflationary epoch:
\eqn\vacen{
V \sim M_{GUT}^4 \sim \left(10^{16}\ GeV\right)^4\ .
}
Here $M_{GUT}$ is the ``unification scale'' in
supersymmetric grand unified models, as predicted
by the running of the observed strong, weak and
electromagnetic couplings above $1\ TeV$ in the
minimal supersymmetric standard model.
If this upper bound is close to the truth,
the vacuum energy can be measured directly
with detectors sensitive to the polarization of the CMBR.

In these scenarios, CMBR anisotropies are generated by
quantum fluctuations of the inflaton and graviton
which ``freeze'' into classical perturbations
at the inflationary Hubble scale, $H \sim 10^{14} GeV$.
These fluctuations are inflated to observable scales
over the 65 e-foldings required to generate
the observed homogeneity and isotropy of the CMBR.
Therefore, inflation acts as both
accelerator and microscope, potentially to energies
11 orders of magnitude higher than those detectable by
terrestrial accelerators.

To this end, a number of authors \refs{\martinbrand, \Niem,
\kempf,\egks,\egkss} have
argued that high-scale physics
can lead to observable effects in the CMBR.
Following this, four of the present authors
\refs{\kkls}\ undertook a systematic
and general study of the effects on CMB observations
of new physics at a scale $M\geq  H$.  Our conclusions
differed from some of the previous work.
Assuming that local, effective field theory applies
at the scale $H$, we argued
that short-distance effects were encapsulated in
irrelevant operators in the inflaton Lagrangian.
Higher-derivative terms lead to
effects which are distinguishable from
corrections to the effective potential,
by modifying a relationship (one of the
inflationary ``consistency conditions''\conscon)
between the temperature and polarization
anisotropies of the CMB.\foot{This
is not the only way to violate these
consistency conditions; scenarios with multiple
scalar fields also lead to a violation
\refs{\multif,\corrconsist,\isoc,\efbond,\langlois}.
However, as argued in \refs{\kkls}, these
deformations require drastic fine tuning of
both the potential energy functional and the
initial conditions for these fields.}
Locality and Lorentz invariance
imply that the effects are of order $(H/M)^2$.

Unlike some of the previous work, the calculations in \refs{\kkls}\
assumed that the quantum state of the inflaton and graviton
at scales close to $H$ was the standard thermal\foot{The standard de Sitter
vacuum is referred to variously as the thermal,
adiabatic, and Bunch-Davies vacuum;
we will use thermal.} vacuum
of de Sitter space.\foot{Of course the inflationary universe
is not quite de Sitter space, as the Hubble constant is changing,
but so long as the ``slow roll'' approximation applies,
dynamics at length scales on the order of or above the
several times the Hubble scale can be approximated as occurring
in de Sitter space.}  The result is that our effects
are smaller than those discussed in the work cited above;
in particular, in \refs{\martinbrand,\egks,\egkss},
the effects are of order $H/M$(the results of
\kempf, however, agree with \kkls).  For this
reason, the results of \refs{\kkls}\ have been criticized as being
``too pessimistic" \refs{\danielsson,\greeneII,\brandenbergerII,
\danielssontwo}.\foot{\refs{\greeneII, \danielsson}
also find that the correction due to the $\alpha$ state
will be sinusoidal in the Hubble parameter,
and so in observed wavenumber at the end of inflation.
This would be a striking signature.}

Generic excitations above the thermal vacuum tend to inflate away --
indeed, this is part of the virtue of inflation.  But
as pointed out in \refs{\danielssontwo}, the quantum states advocated
by \refs{\martinbrand,\egks,\egkss,\danielsson,\greeneII}
as an initial state for inflation
lie in a one-parameter family of
de Sitter-invariant states, constructed and described
in \refs{\cherntag,\schomblond,\mottola,\allen}.
(See \refs{\bms}\ for a particularly clear discussion
of these states).
These states, as we will describe below in
detail, resemble squeezed states or Bogoliubov rotations of
the standard, thermal vacuum.  Relative to this
vacuum, they have excitations at arbitrarily high energies.
In pure de Sitter space,
they make up a family of de Sitter-invariant states $\ket{\alpha}$,
where $\alpha$ is a complex parameter indexing the state;
(thus they are often called the ``$\alpha$-vacua'').
These states appear to result in an
$\alpha$-dependent correction to the standard
formulae for CMB anisotropies.  The effect is of
order $H/M$, where $M$ is the scale of new physics, which is larger
than the $H^2/M^2$ effects discussed in \refs{\kkls, \kempf}.

The states $\ket{\alpha}$
have also made an appearance in work on
the conjectured correspondence between
euclidean conformal field theory and quantum
gravity in de Sitter space \bms.  Changing $\alpha$
is conjectured to correspond
to a marginal deformation of the dual CFT.
This makes them doubly interesting.

Motivated primarily
by the possible utility of these states as inflationary initial
conditions,
we would like to examine them more closely.
We will find that these
states, while perhaps consistent with an absolutely free field
in a fixed background de Sitter space, are inconsistent in
inflationary
cosmology, and indeed are problematic even in exact de Sitter space
once gravitational backreaction is included.

With some optimism about the future sensitivity
of CMB experiments, especially polarization data,
\refs{\kkls}\ estimated that the deviations
from standard predictions would be observable if
the correction -- be it $(H/M)$ or $(H/M)^2$ --
is of order $10^{-1}$.  Since the difference between
scales $M$ detected in each case is then a factor of
a few, this difference may seem academic.  But future gravitational
wave detectors may be more sensitive, because they are not limited
by cosmic variance constraints.  However, uncertainties in
the backgrounds make their ultimate sensitivity hard to
predict (see \refs{\great, \greatb, \greatc}).  If these measurements
were sensitive at the $10^{-4}$ level, the difference in scales probed
by $H/M$ effects versus $H^2/M^2$ would be quite substantial.
So it is experimentally, as well as theoretically, important
to resolve this issue.

\subsec{Summary}

Before continuing, we would like to provide a summary of
our basic arguments.
The usual attitude amongst inflationary theorists is that
any deviation from the thermal vacuum relaxes to
the thermal vacuum within a few e-foldings of inflation.
But in the states $\ket{\alpha}$,
the deviations from the thermal vacuum do not
inflate away, because these deviations extend to
arbitrarily short distances.  As a direct consequence,
the quantum energy-momentum
tensor $\bra{\alpha}T_{\mu\nu}\ket{\alpha}$
diverges.

Of course, the computation of $\tvev$
diverges even in the thermal vacuum, for the same reason,
and with the same coefficient, as in Minkowski space.
A consistent regulator in the thermal vacuum is therefore
also a consistent regulator in flat space; the short-distance
properties of Green functions are the same.
However, $\bra{\alpha}T_{\mu\nu}\ket{\alpha}$
diverges with a coefficient that is $\alpha$-dependent,
as the short-distance behavior
of Green functions computed with $\ket{\alpha}$
differs from those in flat space.

We can choose an ($\alpha$-dependent) regulator, such that
$\bra{\alpha}T_{\mu\nu}\ket{\alpha}$ is finite, but then
the energy-momentum tensor will diverge in Minkowski space, and
in the thermal vacuum of de Sitter.
Of course, the universe today may be approximately de Sitter,
and {\it a priori} could therefore still be in an $\alpha$ state.
However, we will show that values of $\alpha$ which cause
interesting effects on the CMB spectrum produced during inflation
would produce enormously unphysical effects on physics in the world
today.

a clear sign that the states $\ket{\alpha}$ are fundamentally
sick is the response of an inertial detector coupled linearly
to a field in the $\alpha$ state \refs{\unruhd, \bms}.
The detector will equilibrate to a non-thermal
spectrum containing infinite energy excitations, which are unsuppressed
by any Boltzman factor.  This is incompatible with the
``causal patch" description of de Sitter space.  In such coordinates,
physics close to the horizon is best described
by a temperature $T=M_p$ ``stretched horizon'',
where gravitational couplings are
order one \refs{\stretched}.  The stretched horizon
acts as a universal thermalizer.  The system is a finite energy, 
closed, thermal cavity, which
will always equilibrate and thermalize its contents, 
with the thermalization time scale
set in this case by the time it takes an excitation
to fall away from the center of the patch and reach
the stretched horizon.  The non-thermal $\alpha$ state
can either equilibrate with the walls (resulting 
in a thermal vacuum), or is simply inconsistent due
to back-reaction from the infinite energy.
The same class of arguments also show that black
hole radiation must be thermal, despite
the apparent appearance of transplanckian modes
near the horizon.  The thermal nature of black hole
radiation is well-tested in string theory,
and is particularly manifest in studies of the AdS black hole.

This leaves the thermal vacuum, and finite energy excitations
above it, as the only consistent
choice for the initial quantum state of
an interacting field theory in an inflating universe.
Since excitations inflate away and thermalize, the vacuum is the
most generic choice.  Nonetheless,
rather than considering the exact $\alpha$ state, one can consider an excited state
above the thermal vacuum, for example one which is identical to the $\alpha$ state up to
some high scale $E$, and is then populated thermally at higher scales.
This state will have a very large energy if $E$ is large, and hence will
back-react and invalidate the model.  Even if this difficulty is ignored,
or the scale is low enough that it is not a problem, the energy $E$ will
exponentially decrease with time, leading rapidly to a near exact thermal
state.  Therefore, in order to produce interesting effects on the CMB
spectrum, this initial state would have to be fine-tuned to be
important for precisely the 10 e-foldings visible in the fluctuation
spectrum.

The last possibility is to simply cut off
the quantum contributions
to $\tvev_\alpha$ above a fixed, physical scale $M$.  This is the
attitude taken in \refs{\egks}, where $M$ is taken to be the mass scale
where new, unknown physics becomes important.  Of course, any excitations
below the scale $M$ will inflate away within
order $\ln M/H$ e-foldings, in the same manner and for the
same reason more conventional perturbations of the
inflationary initial conditions do.
Therefore, in order to produce effects
that last longer than a few initial e-foldings, one has to postulate
that the new physics at scale $M$ ``pumps" in new modes in an
$\alpha$ configuration for all times during inflation.
As these modes are produced at $M$ and inflate away,
more must be created to prevent the effect from disappearing.
It is possible to choose $M$ and $\alpha$ such that the
extra energy density in the modes below the scale $M$ is small
compared to the curvature during inflation.  However,
this ``pump" must shut itself off at the end of inflation,
as otherwise the extra energy would dominate the
post-inflation universe.  This appears to be
both fine-tuned and a violation of standard Wilsonian decoupling.
Furthermore, the physics above $M$ can not be simply ignored - in
particular, conservation of energy implies that the modes must
be produced out of some energy reservoir, which should
gravitate strongly in much the same way as the original
$\alpha$ state did.  Finally, our understanding of string- and
M-theory, incomplete as it may be, provides no hint of any such
behavior; on the contrary, the stretched horizon (which provides
a cut-off much like this) is explicitly
thermal.

\subsec{Outline}

The remainder of this paper is as follows.  In section 2
we will discuss the states $\ket{\alpha}$ in
eternal de Sitter space.  In the beginning of Section 2 we will
discuss
these states in FRW-type coordinates, and argue
that any consistent regularization will cause the
differences between $\ket{\alpha}$ and the thermal vacuum
to inflate away.  Section
2.2 will discuss the complementary picture of this
argument in the static patch, and argue that the
state $\ket{\alpha}$ corresponds to a non-thermal state which
rapidly thermalizes.  In section 3 we will apply these observations
to an inflationary universe.  We then conclude, and discuss
some details of the thermal character of the $\alpha$ states
in the appendix.

\subsec{Notes on previous work}

The fact that the $\alpha$ states are totally unsuited for use as
initial conditions for inflation
has been noted at least as far back as \mottola,
in which the author mentions that the thermal vacuum is the correct
choice
for a realistic inflation model, for essentially the same reasons we
are
explaining here.  Indeed, many of the arguments
we present here, including the analogy between inflationary
fluctuations and Hawking radiation, informed the
general consensus within the inflation community
that the thermal vacuum is the correct one.\foot{We
would like to thank A. Albrecht and S.-H. Tye for
discussions on this point.}

After this work was completed, we received \banksb\ and \larsen\
which argue that interacting field theory in the $\alpha$ state is
not well defined, due to the appearance of non-local
divergences at one loop.  In addition, \larsen\ reached
conclusions identical to ours on the thermal character of the
$\alpha$ states.  In another related work \goldlowe,
the authors concluded that the states $\ket{\alpha}$
lead to
infinitely large backreaction. The authors then go on to consider a
non-de Sitter-invariant state which leads to corrections
to the CMB spectrum
at order $H/M$.  We discuss such states in \S2, which
arise when one throws away modes of $\ket{\alpha}$ above some
frequency.  As we point out there, states which do not lead to
unacceptable backreaction inflate away within a few e-foldings.
In order to produce a sufficiently flat and homogeneous universe,
inflation needs to have lasted for at least 65 e-foldings.  Therefore,
the state considered in \goldlowe\ would involve energies at least up
to
$e^{65} M$, where $M$ is some high scale.  Such states, if
regulated in a way consistent with either the thermal vacuum
or any $\alpha$ state, would
produce an enormous backreaction which would invalidate the model.
Even
if a non-de Sitter invariant regulator could be found that would
remove
this divergence,
in sec. 3 we point out that a fine tuning is required for
these states to affect CMBR measurements.  Unless the UV scale
considered in \goldlowe\ is tuned precisely to a time corresponding to
the
observable part of inflation, there will be no effect on the spectrum.

\newsec{The states $\ket{\alpha}$ in de Sitter space}

In this section we will argue that the states
$\ket{\alpha}$ are not consistent in a
field theory interacting with gravity in de Sitter space.
It is useful to study this issue from several points of view.

The principle of equivalence states that physics
is invariant under the choice of coordinate system.
Nonetheless, the description may be very different.
We have become used to this fact in studies of black holes.
From the standpoint of an observer at fixed
distance outside a black hole, an infalling object
is blueshifted as it approaches the horizon and
eventually reaches Planckian energies.  A ``stretched
horizon'' \refs{\stretched} can be defined as the place at which modes
which asymptotically have energies equal to the Hawking
temperature are blueshifted to the Planck scale.  This
stretched horizon can be treated as a thermal membrane,
and an infalling observer reaching it will rapidly thermalize
via gravitational and other interactions.
On the other hand, infalling obsevers will fall through
the horizon and not need Planckian physics to describe
this experience.  That these two pictures are equivalent
is demanded by general covariance, and is the
statement of black hole complementarity \refs{\stretched}.

Similar points of view exist in de Sitter space.  Note first
that the causal structure -- the Penrose diagram -- is
the same as that of the AdS-Schwarzchild black hole,
a fact noted by many physicists.  There are static coordinates
which cover the single ``causal diamond'' that an inertial
observer can access in experiments; these are analogous
to Schwarzchild coordinates in a black hole background.
Physics near the horizon can be described by a thermal
``stretched horizon'' just as it is for the
black hole.  There are also coordinates describing
all parts of de Sitter space
which can be influenced by an inertial observer;
these are the analogue of coordinates which can describe
an infalling observer.  The two pictures are
complementary.  The statement in flat coordinates that
an excitation inflates away is related to the
statement in static coordinates that excitations
rapidly thermalize when they approach the causal horizon.
The precise thermality of fluctuations in de Sitter space
follows for the same reason that the Hawking radiation
of a black hole is believed to be thermal.

In the remainder of this section we will analyze
the states $\ket{\alpha}$ from both points of view.

\subsec{$\ket{\alpha}$ in flat coordinates}

Inflation is conventionally described in flat FRW coordinates
(\cf\ \refs{\kolbturner,\lindebook,\reconsrev,\liddlelyth}).
The Robertson-Walker form of the metric is:
\eqn\rwmetric{
ds^2 = -dt^2 + e^{2Ht} d\vec{x}^2.
}
Here $\vec{x}$ is a 3-vector, and $(t,\vec{x})$
take values in all of $\IR^4$.  $H$ is the ``Hubble scale'',
and objects separated by a comoving ($x$-coordinate) spatial distance
larger
than $e^{-t}/H$ are out of causal contact with each other.
These coordinates cover the half of de Sitter space labelled I
in the figure below.

\bigskip
\centerline{\vbox{\hsize=5.0in
\centerline{\psfig{figure=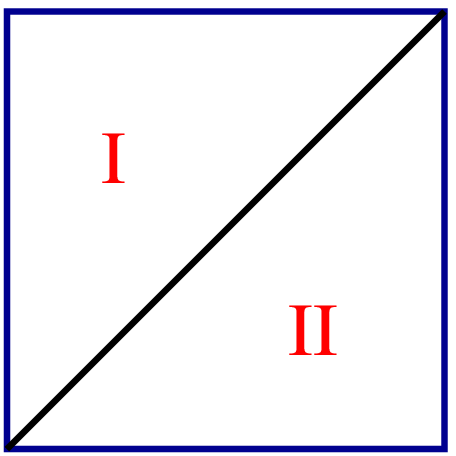}}
\vglue.4in
\noindent {\tenpoint Figure 1. Region I is the spatially flat patch of
de Sitter space
for an observer on the left-hand boundary of the diagram.}}}
\bigskip

Using the transformation
\eqn\transf{
\eta = -\frac{1}{H} e^{-Ht},
}
this same patch can be described in conformally
flat coordinates using the metric
\eqn\confmet{
ds^2 = \frac{1}{H^2\eta^2} \left(
-d\eta^2 + d\vec{x}^2 \right)\ ,
}
where $\eta \in \left(-\infty, 0\right)$.

\subsubsection{Constructing the states}

Let $\phi$ be a free, minimally coupled scalar field
with mass $m$.  This could for example be the inflaton,
for which $m \ll H$.
A complete set of solutions to the
Klein-Gordon equation in conformal
coordinates is \refs{\birrelldavies}:
\eqn\solutions{
\phi(\eta,\vec{x})^{+ \above 0pt -}_k =
\frac{H\pi^{1/2}\eta^{3/2}}{2(2\pi)^{3/2}}
e^{i\vec{k}\cdot \vec{x}}H_\nu^{({2 \above 0pt 1})} (k\eta)\ ,
}
where
\eqn\index{
\nu = \sqrt{\frac{9}{4} - 12\frac{m^2}{H^2}},
}
$k = |\vec{k}|$, and $H_\nu^{({1 \above 0pt 2})}$ are Hankel functions
of the first and second kind.
As $k \to \infty$ (or $\eta\to -\infty$),
\eqn\asymptotics{
H^{( {2 \above 0pt 1} )}(k\eta) \to \left(\frac{2}{\pi k
\eta}\right)
e^{( {- \above 0pt +} ) ik\eta}\ ,
}
which is the standard behavior for a positive (negative) frequency
mode
in flat space.  As expected, the curvature corrections disappear
at large momentum.

The thermal vacuum corresponds to choosing $H^{(2)}$ as the
positive frequency modes.  In other words, one may decompose
the field operator $\phi$ as
\eqn\adiabaticop{
\phi = \sum_k \left( \phi_k^+ a_k +
\phi_k^- a_k^\dagger\right)\ ,
}
where $a_k,a_k^\dagger$ are creation and annihilation
operators satisfying the usual canonical commutation
relations.  The standard thermal vacuum then satisfies
\eqn\usualvacrel{
a_k \ket{{\rm thermal}} = 0\ .
}
The motivation for choosing this vacuum is
that at short distances it looks like the vacuum in flat space,
in accord with the standard intuition that the short-distance
behavior of the theory should be independent of the space-time
curvature.

The thermal vacuum lives in a one-complex parameter
family of states which are invariant under the
isometry group of de Sitter space
\refs{\cherntag,\schomblond,\mottola,\allen}
(see \refs{\bms} for a recent, clear discussion).
This family can be described via Bogoliubov
transformations as follows \refs{\mottola,\allen,\bms}.
For complex parameter $\alpha$,
${\rm Re} (\alpha) < 0$, the new annihilation operators
are \refs{\bms}:
\eqn\alphaop{
a_k^{\alpha} = N_\alpha\left(
a_k - e^{\alpha^\ast} a_k^\dagger
\right)\ ,
}
where
\eqn\normalization{
N_\alpha = \frac{1}{\sqrt{1 - \exp\left(
\alpha + \alpha^\ast\right)}}\ ,
}
and
$$ a_k^\alpha \ket{\alpha} = 0\ . $$
The thermal vacuum is the limit ${\rm Re} (\alpha) \to-\infty$
of these states.  Such a choice corresponds, in the high momentum
limit \asymptotics, to choosing a mixed set of positive and negative
frequency
modes for the creation and annihilation operators.  It is this fact
that
underlies all of the difficulties the states $\ket{\alpha}$ suffer
from.

\subsubsection{Two-point functions}

One may express the various Green functions in terms of the
unordered or Wightman two-point functions (\cf\  \birrelldavies.)
The Wightman functions in the state $\ket{\alpha}$
\eqn\alphawight{
G_\alpha(x,x') =
\bra{\alpha}\phi(x)\phi(x')\ket{\alpha}
}
have a simple expression in terms of the thermal
Wightman function
\eqn\adwight{
G_T = \bra{0}\phi(x)\phi(x')\ket{0}\ .
}

To construct $G_\alpha$, we must continue $G_T$ to
global de Sitter space, defined as the the hyperboloid
\eqn\hyperboloid{
- (X^0)^2 + \sum_{k=1}^{d} (X^k)^2 = \frac{1}{H^2}
}
in $(d+1)$-dimensional Minkowski space $\IR^{1,d}$.
Define the antipodal point of $X(x)$ as $X_A = - X$.
Then $G_\alpha$ can be written as \bms:
\eqn\transf{
\eqalign{
G_\alpha(x,x') & =  |N_\alpha|^2 \left[
G_T(x,x') + \exp\left(\alpha+\alpha^\ast\right)
G_T(x',x) + \right. \cr
&\left. \exp(\alpha^\ast) G_T(x,x'_A)
+ \exp(\alpha) G_T(x_A,x')\right]\ .
}}
The short-distance behavior of $G_\alpha$ is different
for each $\alpha$.  For distances
much smaller than the Hubble scale, $G_T$ looks
like the unordered two-point function in the
Minkowski space vacuum.  At these distances
one can use Minkowski coordinates $t,\vec{x}$, and
the Wightman function approaches:
\eqn\sdwight{
G_T \sim \frac{1}
{\left[(t - t' - i\epsilon)^2 - |\vec{x}-\vec{x}'|^2
\right]^{d/2 - 1}}
}
From this it is easy to see that at short distances:
\eqn\partsofwight{
\eqalign{
\Re\ G_{\alpha} & \sim \frac{1+\exp(\alpha+\alpha^\ast)}
{1-\exp(\alpha+\alpha^\ast)} \Re\ G_T\cr
\Im\ G_\alpha & \sim \Im\ G_T\ .
}}

The Wightman functions can be used to construct the
commutator and anticommutator of $\phi$ and the usual
Green functions (\cf\ \refs{\birrelldavies}.)
The commutator
\eqn\commut{
\bra{\alpha}\left[\phi(t,x), \partial_t\phi(t,x')\right]
\ket{\alpha} =
\lim_{t\to t'}\p_{t'}\left(
G_\alpha(x,x') - G_\alpha(x',x)\right)
}
can be shown using \transf\ to be independent of $\alpha$,
as expected since the free field commutator is a c-number
and is determined uniquely by the operator algebra.
The Feynman propagator
\eqn\feynmanalpha{
\eqalign{
G_{F,\alpha} & = -i\theta(t-t')G_\alpha(x,x') -
i\theta(t'-t)G_\alpha(x',x)\ \cr
&= {1 \over 2}
\left[ \theta(t-t')G_\alpha(x,x') -
\theta(t'-t)G_\alpha(x',x) \right] - {i \over 2}
\left[ G_\alpha(x,x') + G_\alpha(x',x) \right] }}
has different short-distance behavior from $G_{F,0}$, however.
The first term in square brackets is determined by the operator
algebra and is therefore independent of the state,
but the second term--the Hadamard function, which is determined by
the anti-commutator of the fields--is $\alpha$ dependent.

\subsubsection{Particle content of $\ket{\alpha}$.}

%REFERENCING

We can measure the structure of $\ket{\alpha}$
by coupling $\phi$ to
an inertial detector.  Since no stress-energy is
needed to maintain the trajectory of an inertial
observer, any transition the detector makes must
be the result of absorbing quanta of the field $\phi$.

The ``Unruh detector'' \refs{\unruhd}
couples linearly to $\phi$,
\eqn\detectorham{
\delta H = \int d\tau \sqrt{g} m(\tau)
\phi(t(\tau), \vec{x}(\tau))\ .
}
where $m(\tau)$ is some operator measuring the state of a detector,
and $t(\tau),\vec{x}(\tau)$ is the trajectory.
Let us imagine a detector at the origin in inflating
coordinates, so that the coordinate time
and the proper time along the detector's trajectory
are identical.
In this case, the response of an Unruh detector was calculated for
all values of $\alpha$ in \refs{\bms} \foot{The authors of \bms\
computed
this result for 2+1 dimensional de Sitter space.  However, the
computation can be easily
extended to 3+1 dimensions, with the same result for the spectrum of
the Unruh detector.}.
If the initial state of
$\phi$ is $\ket{\alpha}$,
\eqn\responserate{
\dot{P}_{\alpha, i\to j} =
\bra{E_j}m(0)\ket{E_i}
\int_{-\infty}^{\infty} dt
e^{-i\delta E t}G_\alpha(t)
}
is the probability per unit time
for the detector to make a transition from
a state with energy $E_i$ to a state with energy
$E_j = E_i + \delta E$.  We can use general properties
of $G_T$ under translations in imaginary time \refs{\bms}\ to compute
the
ratio for $\delta E \gg \frac{H}{2\pi}$:
\eqn\rateratio{
\CR_{ij} = \frac{\dot{P}_{i\to j}}{\dot{P}_{j\to i}}
= e^{-2\pi \delta E/H} \left|
\frac{1 + e^{\alpha + \pi \delta E/H}}
{1 + e^{\alpha - \pi \delta E/H}} \right|^2 =
f_\alpha(\delta E)\ .
}
This ratio is finite even if we do
not regularize $G_\alpha$.

The thermal vacuum is the limit $\alpha\to -\infty$.
In this state, the detector will equilibrate to
\eqn\densratio{
\frac{\dot{P}_{i\to j}}{\dot{P}_{j\to i}}
= \frac{\rho(E_j)}{\rho(E_i)} = e^{-2\pi \delta E/H}\ ,
}
where $\rho(E)$ is the density of states at energy $E$.
The detector is in equilibrium at temperature
$T_{dS} = \frac{H}{2\pi}$, which is consistent with the usual
statement that the standard vacuum is thermal,
with temperature $T_{dS}$.

If $\alpha \neq -\infty$, however, the detector cannot
reach an equilibrium consistent with the principle of
detailed balance.  Detailed balance combined with
\rateratio\ implies:
\eqn\badratio{
\CR_{ij} = \frac{\rho(E_j)}{\rho(E_i)}\ .
}
But this implies
\eqn\multratio{
\CR_{ij} \CR_{jk} = \CR_{ik}
}
and therefore
\eqn\nocando{
f(\delta E)^2 = f(2\delta E)\ ,
}
which is not true unless $\alpha = -\infty$, and in particular is only
valid for
the Boltzman distribution \densratio.  This does not
indicate that no equilibrium will be achieved (as claimed in \bms).
One can show that a system obeying \rateratio\ does equilibrate,
but the configuration will not
be consistent with detailed balance.
Unitarity, time reversal invariance
of the microphysics, and equipartition imply
detailed balance; our result indicates
that one of these breaks down when coupling a
detector which possesses these properties
to a field in the state $\ket{\alpha}$.

In the thermal vacuum, $\CR_{ij}$
drops exponentially for $\delta E \gg 2\pi H$.
But for {\it finite} $\alpha$ and
$\delta E \gg \pi H$, the ratio asymptotes to
$$ \CR \to e^{2\alpha}\ . $$
For $\delta E \gg \pi H$, there is a
constant probability for the
detector to be excited, for all values of $\delta E$.
This is consistent with $\ket{\alpha}$ having infinite energy.
We will discuss this further in Section 3 and the appendix.

\subsubsection{The quantum stress-energy tensor}

We are particularly interested in the case where $\phi$
is the inflaton.  At distances shorter than $1/H$, it
can be treated as a massless, minimally coupled scalar.
Ignoring potential energy terms, the
expectation value of the stress-energy tensor
can be computed from the anticommutator:
\eqn\unregt{
T^{\alpha,unreg}_{\mu\nu} =
\lim_{x\to x{}'}\left(\p_\mu\p_\nu{}' - \half g_{\mu\nu}
g^{\alpha\beta}\p_\alpha\p_\beta{}'
\right) G^{(+)}_\alpha(x,x')\ .
}
where
\eqn\anticom{
G^{(+)}_\alpha = \bra{\alpha}\left\{\phi(x),\phi(x')\right\}
\ket{\alpha}\ .
}
Because of the short-distance singularity in
$G^{(+)}$, \unregt\ is divergent.  In the limit $x\to x'$,
the last two terms in \transf\ are irrelevant, so that
the divergence in this quantity can be related
to the divergence in the thermal vacuum \mottola\ \allen  :
\eqn\divreln{
T^{\alpha,unreg}_{\mu\nu} =
\frac{1 + \exp\left(\alpha+\alpha^\ast\right)}
{1 - \exp\left(\alpha+\alpha^\ast\right)}
T^{therm,unreg}
}
In the thermal vacuum, sensible regularization procedures
exist, and lead to a finite value for $T_{\mu\nu}$
(\cf \refs{\birrelldavies,\waldbook}).
We now turn to the question of whether a sensible
procedure exists for finite $\alpha$.

At proper distances shorter than the Hubble scale, the background
approaches flat space, and the thermal vacuum approaches
the Minkowski vacuum.  The results of
coupling an Unruh detector to $\ket{\alpha\neq-\infty}$
imply that such states are excitations above the Minkowski vacuum.
At arbitrarily high energy these excitations do {\it not}
fall off in amplitude, so that
the quantum stress energy tensor will be infinite.

We may nonetheless try to regularize $T_{\mu\nu}$. One option,
described in \refs{\egks,\egkss}\ is to compute $T_{\mu\nu}^\alpha$
by expanding $\phi$ into creation and annihilation operators, and
simply discarding the contributions of all modes with proper
momenta higher than some cutoff scale $\Lambda$.  The modes with
lower momenta are taken to contribute to $T_{\mu\nu}$.  As modes
at the scale $\Lambda$ inflate away, new modes carrying
stress-energy must be ``pumped into'' the system at the scale
$\Lambda$ to maintain the deviations from the thermal vacuum.
Prior to their being pumped in, they must not interact with
gravity.  This is unphysical.

A physical procedure for cutting off the theory
would be to choose some particular time
in inflating coordinates, and cut off the excitations
above some momentum scale $\Lambda$.  One must ask what
``excitations'' mean.  The clearest definition,
given our discussion of particle detectors, is excitations
relative to the thermal vacuum.  After all, the
thermal vacuum approaches the standard Minkowski
vacuum at short distances.  In this case the additional
excitations will inflate away in
\eqn\infltime{
N = \ln \frac{\Lambda}{H}
}
Hubble times.

This regulator is not de Sitter invariant, however.
One may instead choose a covariant, de Sitter-invariant
regulator, such as a proper time cutoff or a covariant
point-splitting regulator.  (See
\refs{\birrelldavies,\waldbook} for a
detailed discussion).  Such regulators
subtract the short distance singularity in
the definition of the composite operator
$T_{\mu\nu}$.
But we can see from \divreln\ that
the short distance behavior of the operator product
defining $T_{\mu\nu}$ is different for different
values of $\alpha$.  Similarly, the Unruh detector
calculation leads to a state with a structure
at high energies which depends on $\alpha$.

Therefore, the prescription for cutting off the theory
must depend on $\alpha$ if the vacuum energy
is to remain finite.
%One will
%then subtract from the stress tensor a divergent piece
%with the correct $\alpha$-dependent coefficient
%to yield a finite result.
But each prescription for a
short distance cutoff defines a different field theory.
If we pick a theory for which $\ket{\alpha}$ is well defined,
$\ket{\beta\neq\alpha}$ will not be well defined.
When computing $T_\beta = \bra{\beta}T_{\mu\nu}\ket{\beta}$,
one would subtract the divergence of the OPEs
in the vacuum $\ket{\alpha}$.  From \divreln\ it
immediately follows that $T_\beta$ is infinite.

While one might fantasize about a theory consistent
with a vacuum $\ket{\alpha}$ in eternal de Sitter space,
in the standard inflationary scenarios inflation
ends and the universe eventually relaxes to a state
with the puzzlingly small cosmological constant that
we observe.  A theory which describes the
world should be consistent
with the Minkowski space vacuum at distances
significantly shorter than the {\it present} Hubble scale.
This theory must then be defined with a regulator
which yields a zero-energy Minkowski vacuum in flat space.
Therefore, either the state of the inflaton
during inflation is a finite excitation
above the thermal vacuum, and relaxes to it
in a few e-foldings, or the short-distance structure
of the theory must change with $H$ during inflation.
This latter scenario violates decoupling;
the theory is nonlocal at the scale $H$.
At this point we must simply throw up our
hands.  This kind of nonlocality does not appear
in string theory, which is local down at least to the string scale
$M_s$.
So long as the string scale is higher energy than the Hubble scale,
effective field theory is valid between these
scales and decoupling will hold.  If one allows $M_s \leq H$,
as in \refs{\brandenbergerII},
the de Sitter temperature is greater than or equal to the Hagedorn
temperature.  In this case
the standard picture of inflation completely breaks down
and the calculations in all works in question
are invalid.\foot{Given the mysteries of the cosmological
constant and the vacuum energy, one might try to subtract off
the varying vacuum energy by adjusting the inflaton potential.
But the time at which the $\alpha$ state decays to the thermal
vacuum is not tied to the vev of the inflaton, so it is very
unclear how to implement this.  This issue is related to the
fine-tuning issues discussed elsewhere in this work.}

\subsec{The vacuum in the static patch}

\subsubsection{Coordinate system}

The shaded region in Figure 2 can be described by static coordinates,
which are the analog of Schwarzchild coordinates in a
black hole background.  The metric is:
\eqn\static{
ds^2 = -(1 - (r/R)^2) dt_s^2 + (1-(r/R)^2)^{-1} dr^2 + r^2
d\Omega_2^2,
}
where $R = 1/H$.

\bigskip
\centerline{\vbox{\hsize=5.0in
\centerline{\psfig{figure=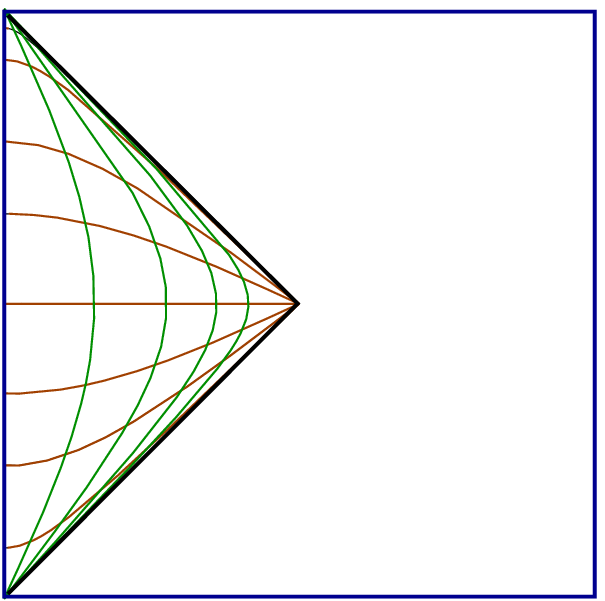}}
\vglue.4in
\noindent {\tenpoint Figure 2.
The static patch of de Sitter, with lines of constant
time and radial coordinate depicted.}
}}
\bigskip

A generalization of these coordinates
also exists when the Hubble parameter changes with time
\mattlennylisa,
so that we can sensibly discuss the physics of inflation
from the point of view of a single inertial observer.
This is the analog in de Sitter space
of studying physics from the point of view of an observer
outside the black hole.

Eq. \static\ describes a spherically symmetric geometry,
with the origin surrounded by a horizon at $r = R$.
The near-horizon limit of \static\ is simply Rindler
space; the horizon behaves much
like the horizon of a Schwarzschild black hole.
At $r= 0$, the time coordinate $t_s$
is identical to the time coordinate in the inflating
and global coordinates.  Therefore, according to
an inertial observer at $r=0$, the vacuum has temperature $T_{dS}$.

For an observer at fixed $r\neq 0$,
the local temperature blueshifts as $r$ approaches $R$,
and becomes infinite at the horizon.  One may cut off the space
by introducing a stretched horizon (rendering the entropy finite),
as advocated in
\refs{\stretched} for black holes.  The stretched
horizon can be defined, for example,
as the surface a Planck length away from the true
horizon.  It can be thought of
as a hot membrane with $T = M_p$.

De Sitter space can therefore be described as a closed
system with thermal walls, a ``hot tin can."  Such a
system will always thermalize, in a time determined by
the initial state.  This is a very general statement,
which does not depend on the details of the interactions
or the spectrum of the hamiltonian.  The only way it can be
avoided is if the system is exactly free, so that it does
not interact with the hot walls at all, or if it contains
infinite energy densities, in which case the thermalization
time can be taken to infinity.

Neither possibility
is consistent with a theory coupled to gravity.
In such theories, if an observer ever
detects a finite-energy fluctuation originating from near the past
horizon,
it must have had
a very large energy where it originated from, which was redshifted
away
during the propagation away from the horizon.
Such a fluctuation will rapidly thermalize in the presence of gravity,
which becomes very strong at high energies.
In other words, take any finite energy fluctuation
detected by our observer and boost it back to where it originated
from.
In a rest frame close to the horizon the fluctuation will be very high
energy, therefore very short wavelength.  Interactions with other such
high energy quanta will very quickly produce a thermal distribution.

\subsubsection{Thermalization time}

Consider a state with some excitations in
the vicinity of the origin, in the coordinates
\static.  If the excitation spectrum is thermal
with $T = T_{dS}$, the excitations will be
in equilibrium with the radiation from the
stretched horizon.

The states $\ket{\alpha}$ are {\it not} thermal (see \rateratio;
$t_s$ and $t$ are identical at $r = \vec{x} = 0$.)  In particular,
the detector coupled to a field in the state $\ket{\alpha}$
is excited with finite and constant probability per unit time
for arbitrarily large $\delta E$, due to the absorbtion
of highly energetic particles.
%In a free theory,
These energetic particles will generally
quickly exit the region of the origin
and arrive at the horizon.  In the case
at hand we can certainly describe them
as relativistic.
The inflaton will have mass $m\ll H$, while the excitations
we describe have energy much greater than $H$, so we can
ignore their rest mass.

We can thus approximate the time for this
excitation to travel from the origin to the stretched horizon,
as the time along a massless geodesic:
\eqn\timeoflfight{
t_s = \int_0^{R - L_p} (1 - (r/R)^2)^{-1} dr =
{R \over 2} \ln{({2 R - L_p \over L_p})}\ .
}
In other words, it will reach the horizon
in $N = Ht = (1/2) \ln (2 M_p / H) < 7$ e-foldings of inflation, using
the most optimistic (for generating effects on the CMB)
values of $H$ and the cutoff.  Note that this is roughly consistent
with the estimate \infltime\ for an excitation to
inflate away in flat coordinates.
(Massive particles will take longer to reach the horizon:  they
will accelerate exponentially away from the origin, until
they reach a distance comparable to the Hubble
length.)

When the particle arrives at the stretched horizon,
it will be blueshifted to energies greater than $M_p$
relative to an observer in the center.
Since the temperature of the stretched horizon is $M_p$,
gravitational interactions will be
of order unity, and the excitation will immediately thermalize.
Any non-gravitational interactions will generically
only thermalize the particle more quickly.

If the state has an infinite number of arbitrarily
high excitations, as is the case with
$\ket{\alpha}$ {\it sans}\ regularization, this argument will fail.
However, in this case backreaction will destroy the
de Sitter geometry.

On the other hand, if the boundary conditions at
the horizon are such that
there is a non-zero flux of particles coming in or out,
the naive argument could also fail.  In this case, however, there will
be
constant collisions at the stretched horizon
between the incoming and outgoing fluxes, or between the
flux and the thermal background, with energies of order the Planck
mass.
This also will take the model out of perturbative control,
as the back-reaction at the horizon will be large.

To avoid these problems, one could cut off the $\alpha$ state.
This will generically break the de Sitter invariance,
leaving a state of the type described above (so that the state will
thermalize in $\sim 10$ e-foldings).
%A de Sitter invariant
%regulator, such as Pauli-Villars, will render
%the theory non-unitary (SAY MORE OR LESS HERE??).

\newsec{Inflation and $\ket{\alpha}$}

The utility of inflation stems from its
predictive power.  The common lore %n
is that ``anything will inflate
away," leaving a flat and homogeneous spacetime at the
end of inflation, which matches our observations.
But of course the universe is not perfectly
homogeneous, and in fact the inclusion of
quantum fluctuations in the theory provides a beautiful
explanation for the CMBR anisotropy and the eventual
formation of large-scale structure.

Observations {\it do} force a certain degree of fine-tuning
in the effective field theory of the inflaton sector.  In particular,
the
potential for the inflaton field
must be quite flat, to allow the universe to expand superluminally %n
(typically 60--70 e-foldings suffice), and to explain the
small CMBR anisotropies $\frac{\delta\rho}{\rho}\sim 10^{-5}$.
The predictivity of inflation then comes from the
fact that, given this assumption,
the initial conditions are completely irrelevant to the spectrum we
observe today, because initial inhomogeneities
are ``inflated away''.  This solves the so-called
horizon and flatness problems (but see \mattlennylisa\
for a discussion of the genericity of inflationary
initial conditions).

One can easily conjure
models which satisfy the slow roll
conditions and the usual assumptions of effective field theory,
and yet which predict unusual features in the
fluctuation spectrum, detectable by the
next generation of precision CMBR experiments.
For example, one
may add a field which
couples to the inflaton and which becomes light ($m\ll H$)
between 70 and 50 e-foldings before the end of
inflation \refs{\wimpz,\kkls}.
This will lead to a noticeable feature in the
spectrum of CMBR anisotropies.
A suitable choice of parameters takes this feature to
the threshold
of observability.  One
can even add a whole series of such fields, providing potentially
infinite freedom to manipulate the predictions.  Such models,
while perhaps interesting in principle, are highly non-generic
in that they require a degree of fine-tuning far above and beyond
what is necessary merely to match the observed
scale invariant spectrum.
However, these models have the virtue that they are
sensible effective
field theories.  Their shortcoming
is of course that they are
even more unnatural than the simplest
inflationary models.  In general, field theorists
prefer not to give up naturalness unless they are forced to
by the data.

Our discussion in the previous section indicates that
the putative distortions in the CMBR spectrum which
arise from considering the inflaton in a state
$\ket{\alpha\neq -\infty}$ is as bad, or potentially
worse, than the other fine tuning issues.  Unregulated,
the states are pathological. If we impose the simplest cutoffs,
the states relax quickly to the thermal vacuum.
We can get an observable effect in the CMBR only if
we fine tune the initial conditions so that inflation begins
precisely at 65 e-foldings, so that the distortions occur
within the first ten e-foldings of inflation before
they inflate away.  But in general there is no reason
that inflation should not last for 70 or 80 or 100 e-foldings.
Furthermore, all conventional models of inflation
assume the initial state is inhomogeneous to some degree.
If inflation began precisely 65 e-foldings ago,  the
CMB spectrum would be distorted simply by the fact
that the initial inhomogeneities in the inflaton
would not have had time to be ironed out by the
expansion.

We should note that while CMBR anisotropy measurements
are sensitive to
effects generated between 55-65 e-foldings before
the end of inflation,
gravitational wave detectors are sensitive
to fluctuations generated much later.
Therefore, even finely-tuned states will not affect
these latter experiments.

Part of our argument that $\ket{\alpha}$ is sick
rested on the observation that the
universe today is nearly flat, and that the vacuum state
is close to the Minkowski vacuum.  In this section we
will provide further arguments to show that the states
$\ket{\alpha}$ are pathological in a realistic cosmology.

\subsec{The causal structure of $\ket{\alpha}$}

A worrysome
aspect of the $\ket{\alpha}$ states
is the presence of singularities
in the Green functions at spacelike separated points
(in particular, the Green functions
diverge when the points are antipodal; see Figure 3).

\bigskip
\centerline{\vbox{\hsize=3.0in
\centerline{\psfig{figure=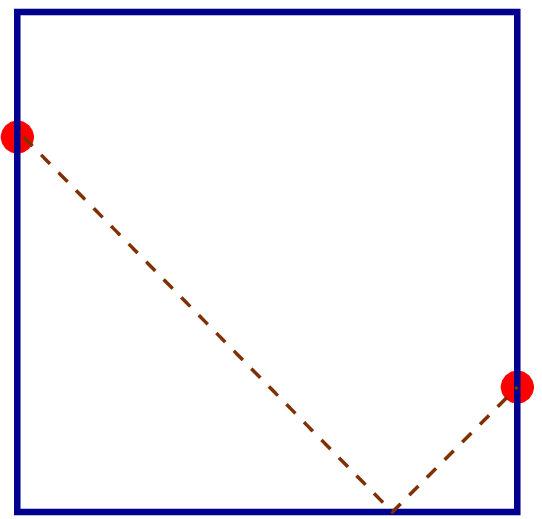}}
\vglue.4in
{\tenpoint Figure 3.
Two antipodal points in de Sitter.}
}}
\bigskip

The retarded and advanced propagators do not suffer from this disease,
as they depend only on the commutator of fields and as
such are determined purely by the operator algebra.  For this
reason we do not expect the theory to be acausal,
despite the presence of spacelike singularities.  Furthermore,
in eternal de Sitter, the causal patch and its antipodal
image only overlap at $t = \pm \infty$ (in other words, the
antipodal point is always ``behind the horizon").  However,
any perturbation to exact de Sitter elongates the Penrose
diagram, which brings some antipodal points into the past
lightcone of the origin at finite time (see Figure 4).

\bigskip
\centerline{\vbox{\hsize=5.0in
\centerline{\psfig{figure=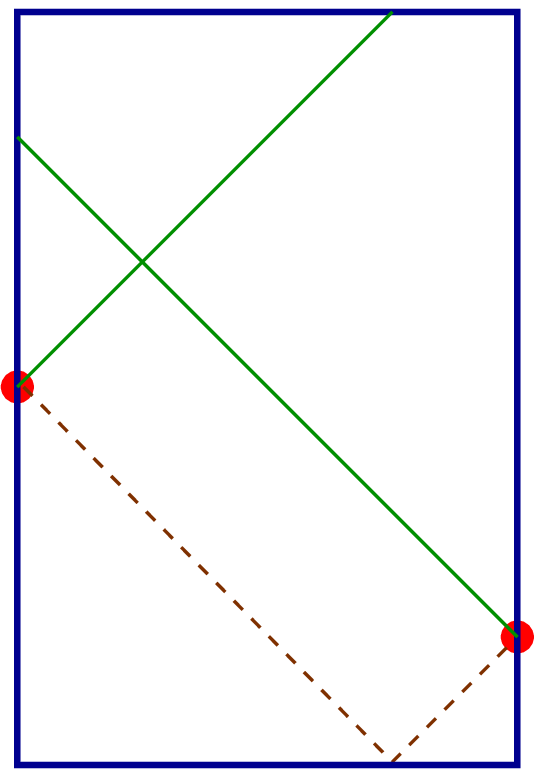}}
\vglue.6in
\noindent {\tenpoint Figure 4.
The Penrose diagram for a de Sitter space
with some matter present,
or for one which transitions from inflation (the bottom part of the
diagram)
to a final, larger de Sitter.
In such a space the future lightcones of some of the
antipodal points in the inflating patch can intersect.}
}}
\bigskip

Although this apparently does not render the theory
acausal, it is disturbing, and probably indicates in
yet another way the pathological nature of
the $\ket{\alpha}$ states.

%\lref\wald{ R.~M.~Wald,
%``General Relativity,''
%{\it  Chicago, Usa: Univ. Pr. ( 1984) 491p}. }

\lref\Radzikowski{ M.~J.~Radzikowski,
``Micro-Local Approach To The Hadamard Condition In Quantum
Field Theory On Curved Space-Time,''
Commun.\ Math.\ Phys.\  {\bf 179}, 529 (1996).
%%CITATION = CMPHA,179,529;%%
}

Wald has argued \refs{\waldbook,\Radzikowski} that a vacuum in a QFT
with
states satisfying a standard positive norm condition, and the
condition that the short distance behavior of the Green functions
match flat space, is sufficient to guarantee the absence of such
spacelike separated singularities.  He therefore proposed as a
criterion for curved space quantum field theory that the short
distance behavior should match to flat space.  The $\alpha$ states
are examples of states which fail to meet this criteria.

%n
\subsec{Inflationary fluctuations, thermal vacuum and ``other" Hubble
volumes}

When the thermal vacuum prescription is adopted, the standard
causal dynamics for generation of inflationary fluctuations
produces the seeds for large scale structure, and sources the CMB
anisotropies. We have reviewed this classic result
\mukh\starob\hawk\guth\bst\alinsf\ in our earlier work \kkls. Here
we underscore the crucial role of the thermal vacuum in particular
for the generation of fluctuations.

The inflationary density fluctuations come from imprinting the quantum
inflaton
fluctuations on the background geometry. They are generated by the
process of inflationary stretching of the fluctuations to the scales
larger than the apparent horizon, where the fluctuations freeze, and
remain decoupled until horizon reentry much after inflation.
However the universe which we presently experience consists of a huge
number of domains which were Hubble size at the time of inflation, and
therefore causally disconnected. If we take the scale of
inflation to be $H \sim 10^{14} GeV$, then the ratio of the observable
volume of the
universe today to the volume then was
\eqn\nodoms{n_d \sim {H^{-3}_{now} \over H^{-3}_{inflation}}
{a^3(inflation) \over a^3(now)} > 10^{78}.
}
While inflation evicted initial inhomogeneities
from these domains, naively one may worry that
the subsequent production of density fluctuations could reproduce
inhomogeneities which are larger than those we observe,
$\delta \rho/\rho \sim 10^{-5}$, because the inflaton
fluctuated independently in each of the many Hubble regions.

Closer scrutiny reveals that this is not the case, and the
thermal vacuum prescription plays a key role in this.
Namely, the equivalence principle guarantees that
the local physics in each Hubble domain is
identical. However, the Hubble domains which are not centered
around our observer are boosted relative to her, and thus in terms
of her reference frame the phenomena in these domains occur
at much higher proper energies, in exactly the same manner as we have
discussed in Sec. 2.1. Thus, in the thermal vacuum, the fluctuations
originating in
distant Hubble domains which may have finite energy in the center
of the inflating patch will have a very small amplitude, because they
had thermalized before reaching the observer.  A vacuum state
containing
high energy excitations which are not exponentially suppressed,
however,
would not be protected by this mechanism and as such would be expected
to
produce large inhomogeneities.
It is possible to arrange for some nonthermal effects
to survive until the end of inflaton, if it is not eternal. However, as
we have stressed
above, this is just another guise of fine tuning, and therefore of
little
predictive value.

%n
\lref\shiwas{ G.~Shiu and I.~Wasserman, ``On the signature of
short distance scale in the cosmic microwave  background,'' Phys.\
Lett.\ {\bf B536} (2002) 1, hep-th/0203113.
%%CITATION = HEP-TH 0203113;%%
}

\subsec{The $\alpha$ state in the post-inflationary universe}

Let us assume for a moment that the in-principle difficulties the
state $\ket{\alpha}$ can be resolved with a suitable de Sitter
invariant regulator, so that the state does not relax to the
thermal vacuum.

In \S2.1\ we reminded the reader that correlation functions in the
state $\ket{\alpha}$ have short-distance behavior which differs
from the behavior in the thermal vacuum or in flat space.  In
order to make the stress tensor finite in the state
$\ket{\alpha}$, we must cut off the theory in a {\it different}\
way than we would in flat space.  However, the epoch of inflation
must have come to an end; the universe today has a tiny
cosmological constant and nearly zero curvature. This causes a
serious problem with basic (\eg\ atomic) physics.

The authors of
\refs{\egks,\egkss,\danielsson,\greeneII,\danielssontwo}\
considered as a state for the inflaton $\ket{\alpha}$, with,
for example, $e^{\alpha} = H_i / M \sim 10^{-5}$ where $H_i$ is the Hubble
constant during inflation.
  Imagine a hypothetical Unruh detector
in this background.  As we demonstrate in the appendix, the
population densities at equilibrium are not thermal, and in fact
tend to a constant at high
energies.  The deviation from thermality becomes significant
for energies $E$ satisfying $E \gg H \alpha$.
During inflation,  $H_i \alpha = H_i \ln (M/H_i)$, which is large,
and hence if any sensible regulation of the $\alpha$ vacuum were
possible, the distortion from the thermal spectrum below this scale
will not be
serious. Indeed, it is precisely this small distortion which is
argued to lead to interesting (order $H / M$) corrections to the
CMB spectrum, in conflict with the effective field theory analysis
of \kkls.

However, let us consider the universe today, with $H_t$ being the
Hubble constant today.  The question is, what is the state of the
inflaton today?  Unless we assume that the short-distance
structure is non-local on a scale set by the Hubble constant of
the universe, so that $\alpha$ is $H$ dependent, it will still be
in the same $\alpha$ state today that it was during
inflation\foot{Of course, it has been proposed by many authors
that a solution to the cosmological constant problem requires
mixing between the scale of the present horizon and the
fundamental UV scales of the theory.  However, the best guess for
the effects of this mixing during inflation would be that they
make the number of degrees of freedom in an inflating patch finite
\refs{\hogan, \bankscosmo, \fischler}.  The effects on
observations will be much smaller than the effects discussed here
\refs{\stevescott}.}.

In this case basic atomic physics is impossible!  The condition
\eqn\condforbadness{
E / H_t \gg \alpha
}
implies that
\eqn\particularbadness{
E \gg \alpha \times 10^{-31} eV\ ,
}
(recall that $\exp(\alpha) \sim 10^{-5}$ to produce interesting
effects on the CMB spectrum). Consider, for example, a transition
from the ground state to any excited state with an energy greater
than $10^{-31}\ eV$.  The rate will be approximately $e^{2 \alpha}
\sim 10^{-10}$, which is {\it independent} of the energy of the
excited state. Far worse, however, the population density of states in
this equilibrium will be roughly constant, at least up to an
energy of order $e^{2 \alpha} M \sim 10^9$ GeV.  This is obviously
a disaster. In particular, a hydrogen atom coupled to fields in
this vacuum will be ionized with probability equal to unity.

To avoid this, one should require $e^{\alpha} = H_t/M_p$.
However, this gives effects on the CMB spectrum of one part in
$10^{60}$, which are obviously totally unobservable.  The only
other option is to require that $\alpha$ change with time as the
Hubble constant changes, for example so as to always satisfy
$e^{\alpha(t)} = H(t)/M_p$. As we have argued, this is a massive
violation of decoupling, and implies that the theory is nonlocal
on the scale $H(t)$.

\newsec{Conclusions}

In this note we have reviewed the choice of ground state in de
Sitter space and in an inflating universe with a slowly rolling
inflaton. It has been suggested that de Sitter symmetries allow
one to pick any de Sitter-invariant state $\ket{\alpha}$.  We have
argued, however, that any $\alpha\neq -\infty$ is only consistent
if the field theory is completely free, and backreaction is
ignored.
%Once
%interactions are taken into account, internal
%consistency considerations based on conventional effective field
%theory arguments at low energies lead to inconsistences in the
%choices of all de Sitter invariant vacua but the thermal one.
All other excitations inflate away or thermalize rapidly.  This
is similar to the situation familiar in the black hole physics,
where the thermal vacuum is the choice consistent both with
effective field theory and holography.

From the standpoint of inflationary cosmology, this choice then
implies that for computations of density perturbations generated
during inflation, the correct procedure is to use local effective
field theory in the thermal vacuum, as in \refs{\kkls}. The
high-energy corrections to inflationary density fluctuations will
come in the form of an expansion in even powers of $H/M$, as has
been discussed in \kkls. We {\it are} assuming that the scale at
which the theory becomes nonlocal is shorter than the Hubble
scale, but if not, the scenario will be rather different from
standard inflation and no calculations performed to date will be
applicable.

We close with a few additional comments.

\subsec{Higher-dimension operators and inflationary perturbations}

Within the context of effective field theory, Shiu and Wasserman
\shiwas\ have recently pointed out a class of higher-derivative
terms which lead to potentially larger corrections
%to the standard formula
%%
%\eqn\theusual{
%    \delta\phi^2(k=H) = \frac{H^2}{4\pi^2}
%}
%% which are potentially larger
than those described in \kkls.

In general, terms of the form
\eqn\polynterms{
\delta_{2n}\CL = \frac{1}{\Lambda^{4n - 4}}
\left(\partial\phi\right)^{2n}\ ,
}
will appear in the inflaton effective action, with $\Lambda$ the
scale of the physics which controls such operators. The even
powers are required by Lorentz invariance. Then during inflation,
substituting $\phi = \phi_0(t) + {\tilde\phi}$ into \polynterms,
where $\phi_0(t)$ is the homogenous mode of the inflaton, using
the slow roll conditions,
\eqn\phidot{
\frac{(\dot \phi)^2}{\Lambda^4} = 2\epsilon
\left(\frac{m_4^2}{\Lambda^2}\right)
\left(\frac{H^2}{\Lambda^2}\right),
}
we find htat $\delta_4 \CL$ leads to a wavefunction
renormalization
of the form:
\eqn\renorm{
\delta \CL = \left(\frac{m_4^2}{\Lambda^2}\right)
\left(\frac{H^2}{\Lambda^2}\right)
\left(\p {\tilde\phi}\right)^2\ .
}
This is consistent with \kkls, in the sense that it is an even
power of $\frac{H}{\Lambda}$. The point of \refs{\shiwas}\ is that
the {\it coefficient} is proportional to
$\frac{m_4^2}{\Lambda^2}$, which may be large. Along the lines of
the arguments in \kkls, this would yield observably large
corrections to the consistency condition for $n_T$ if $\Lambda\sim
M_{GUT}$. For example, the Kaluza-Klein modes in Ho\v{r}ava-Witten
models consistent with grand unification would generically have
observable effects on the spectrum of CMBR anisotropies. This is
an exciting prospect.

Unfortunately, we believe that in realistic theories one should
not equate $\Lambda$ in \polynterms\ to the fundamental scale of
the theory.  For example, in M-theory compactifications, the
inflaton $\phi$ is naturally a compactification modulus. Following
the line of reasoning in \refs{\tomcosm,\tomcosmrev}, we find that
$\Lambda\sim\sqrt{m_{11}m_4}$, as follows.

The inflaton naturally arises as a dimensionless scalar $\psi$
because it is a component of the eleven-dimensional metric. In the
four-dimensional effective action, the kinetic term and the terms
\polynterms\ will take the form:
\eqn\genkinet{
\CL_{kin} = m_4^2 \left(\partial\psi\right)^2
{\cal F}\left(\frac{(\partial\psi)^2}{m_{11}^2}\right)\ ,
}
where ${\cal F}(0) = 1$.  Here $m_{11}$ is the eleven-dimensional
Planck scale.  The factor of $m_4^2$ in front arises from the
large volume of the compactification, in the same way that it
appears in front of the four-dimensional Einstein action. Defining
$\phi = m_4 \psi$, to normalize $\phi$ canonically in the
effective 4d theory, \genkinet\ becomes
\eqn\genkinetII{
\CL_{kin} = \left(\p\phi\right)^2
{\cal F} \left(\frac{(\p\phi)^2}{m_4^2 m_{11}^2}\right)\ .
}
As we claimed, $\Lambda\sim\sqrt{m_{11}m_4}$.  We intend to
investigate this further in various contexts such as brane-world
scenarios and warped compactifications.

\vskip1cm \centerline{\bf{Acknowledgements}}
%%%
It is a pleasure to thank A. Albrecht, C. Burgess, B. Greene,
F. Larsen, A. Linde, J.
Maldacena, D. Spergel,A. Strominger, S.-H. Tye,  R. Wald, and E. Witten for
discussions.  This work is supported in part by  NSF grant
PHY-9870115 and by the Stanford Institute for Theoretical Physics.
Albion Lawrence was also supported in part by the DOE under
contract DE-AC03-76SF00515. Matthew Kleban is the Mellam Family
Foundation Graduate Fellow.
\newsec{Appendix A.}
As we discussed in Section 2, the probability per unit time $\dot{P}_{\alpha, i\to j}$ for
an excitation of an Unruh detector, coupled to a field in an $\alpha$ state,
to transition from a state $E_i$ to a state $E_j$, can be expressed
\eqn\responserateagain{
\dot{P}_{ \alpha , i\to j} =
|  \bra{E_j}m(0)\ket{E_i}|^2
\int_{-\infty}^{\infty} dt
e^{-i\delta E t}G_\alpha(t) \equiv |\bra{E_j}m(0)\ket{E_i}|^2 {\cal F}_\alpha(\Delta E).
}
Here ${\cal F}_\alpha(E)$ is the ``response function" of the detector.
It depends only on the state of the scalar, not on any internal details
of the detector, and hence is a useful diagnostic for the nature of the
state.  Using general properties of the Green function under imaginary time shifts,
one obtains (at least in 2+1 \bms\ and 3+1 dimensions) the ratio
\eqn\rateratioagain{
\CR_{ij} = \frac{\dot{P}_{i\to j}}{\dot{P}_{j\to i}}
= e^{-2\pi \Delta E/H} \left|
\frac{1 + e^{\alpha + \pi \Delta E/H}}
{1 + e^{\alpha - \pi \Delta E/H}} \right|^2.
}
In a system which equilibrates in accord with the principle of detailed balance,
this ratio suffices to determine the relative population densities.
However, as we argued in Section 2,
the equilibrium reached by a system in contact with an $\alpha$ state will not be
consistent with detailed balance.
In this appendix, we will compute $\cal F_\alpha$ exactly for the case of a massless,
conformally coupled scalar in 3+1 de Sitter
(which is identical for these purposes to a minimally
coupled scalar with mass $\sqrt{2} H$).

For such a scalar, the positive frequency Green function in the thermal vacuum is
\eqn\confscalar{
G_0(x, x') = {- H^2 \over 4 \pi^2}{\eta \eta'  \over \left[(\Delta \eta)^2 -
| \Delta {\bf x} |^2 \right]},
}
where $\eta = - e^{-H t} / H$.  Using the identity $\pi^2 \sin^{-2}(\pi x) =
\sum_{n = -\infty}^\infty (x - n)^{-2}$, and setting $ \Delta {\bf x} = 0$,
\eqn\response{\eqalign{
{\cal F}_0(E)  =& {H^2 \over 32 \pi^2} \int_{-\infty}^\infty d (t + t')
\int_{-\infty}^\infty d \Delta t { e^{- i E \Delta t} \over
\sin^2(i H \Delta t / 2 + \epsilon)}
\cr  = & {H^2 \over 32 \pi^4} \int_{-\infty}^\infty d (t + t')\sum_{n = -\infty}^\infty
\int_{-\infty}^\infty d \Delta t  {e^{- i E \Delta t} \over
(i H \Delta t / 2 \pi + \epsilon - n)^2},
}}
where we have included an $\epsilon$ prescription to avoid the pole at $\Delta t = 0$.
This integral can now be computed either by completing the contour above
or below the real axis.  The residues of the integrand are
$-(4 \pi^2 i E/H^2) e^{-2 \pi n E/H}$.  If $\epsilon >0$ (which is the correct
prescription) and we complete the contour below the real axis,
we miss the pole near the origin, and obtain a sum of the form
\eqn\responseb{
\dot{{\cal F}}_0(E) = (E / 2 \pi) \sum_{n=1}^\infty e^{-2 \pi n E/H} =
(E / 2 \pi)  ( e^{2 \pi  E/H} -1)^{-1}.
}
This is a Planck distribution, which is consistent with the thermal
character of the standard de Sitter vacuum.  On the other hand,
had we chosen the incorrect $\epsilon$ prescription, we would have picked up
the pole at the origin, obtaining
\eqn\responsec{
\dot{{\cal F}}_{-}(E) =
\sum_{n=0}^\infty e^{-2 \pi n E/H} =
(E / 2 \pi)  (1 - e^{-2 \pi  E/H})^{-1}.
}
This is an unphysical distribution, which blows up at large energies.

In an $\alpha$ state, the Green function can be expressed as \transf
\eqn\transfb{
\eqalign{
G_\alpha(x,x') & =  |N_\alpha|^2 \left[
G_0(x,x') + \exp\left(\alpha+\alpha^\ast\right)
G_0(x',x) + \right. \cr
&\left. \exp(\alpha^\ast) G_0(x,x'_A)
+ \exp(\alpha) G_0(x_A,x')\right]\ .
}}
The second term involves
the wrong ordering of $x, x'$, so that $\Delta t \rightarrow -\Delta t$, which
is equivalent to changing the sign of $\epsilon$, and therefore contributes
a term proportional
to \responsec.  The terms involving the antipodal
points can be evaluated using the identity
$G_0(x, x'_A) = G_0(x_A, x') = G_0(t - i \pi/H),$
for $x, x'$ at the origin.  From these we obtain
\eqn\antipode{\eqalign{
& \int_{-\infty}^{\infty} d\Delta t {e^{- i E \Delta t} \over
\sin^2\left( i H \Delta (t - i \pi/H)/2 \right)} = {1 \over \pi^2}
\sum_{-\infty}^{\infty}
\int_{-\infty}^{\infty} d \Delta t
{e^{- i E \Delta t} \over (i H \Delta t /(2 \pi) + 1/2 - n)^2} \cr
& = (2 E/\pi)  e^{ \pi E /H} (e^{2 \pi E/H} - 1)^{-1}}.}

Plugging everything back into \transfb, we obtain the response function
for arbitrary $\alpha$:
\eqn\responsed{\eqalign{
\dot{{\cal F}}_{\alpha}(E) =& N_\alpha^2 {(E / 2 \pi) \over e^{2 \pi E/H} - 1}
\left[
1 +
e^{\alpha+\alpha^\ast + 2 \pi E/H} + e^{\alpha + \pi E/H}
+ e^{\alpha^\ast + \pi E/H} \right] \cr
=& N_\alpha^2 \left| 1 + e^{\alpha + \pi E/H} \right|^2
{(E / 2 \pi) \over e^{2 \pi E/H} - 1} \cr =&
N_\alpha^2 \left| 1 + e^{\alpha + \pi E/H} \right|^2 {\cal F}_0(E),
}}
in exact agreement with \rateratioagain, and with the result obtained in
\bms\ for the case of 2+1 dimensional de Sitter.

As we discussed in Section 2, such a distribution is not consistent
with detailed balance, and the density of states of a detector
coupled to it appears
to depend on the details of the detector (contrary to the thermal case).
However, we can make a general statement (assuming the matrix elements
$\bra{E_j}m(0)\ket{E_i}$ are not zero; that is, that the detector is
capable of being excited to these energies) about the equilibrium that a system
coupled to a scalar in the $\alpha$ state will reach.  The general condition
for equilibrium is
\eqn\equil{
\int_{E_0} \left( \rho(E') \dot{P}_{E'\rightarrow E}
- \rho(E) \dot{P}_{E \rightarrow E'} \right) d E' = 0.}
For an ordinary Planck distribution, $\dot P$ dies exponentially
at large, positive energies.  The $\alpha$ distribution \responsed, however,
grows linearly with energy, both for large positive and
negative $\Delta E$.  Therefore, the high energy density of states $\rho(E')$
in the first term in \equil\ must tend to a constant at large energies, in order
to cancel the second term, which will pick up an infinite contribution in the UV.
In other words, the high-energy population densities tend to a constant, rather
than falling off exponentially with the energy as would be the case in a thermal distribution.
\listrefs
\end